\documentclass[sigconf,nonacm]{acmart}
\AtBeginDocument{%
  }

\setcopyright{none}
\copyrightyear{2026}
\acmYear{2026}
\acmDOI{}
\acmConference[AgenticSE @ KDD '26]{Workshop on Agentic Software Engineering (SE 3.0)}{August 10, 2026}{Jeju, Korea}
\acmBooktitle{Workshop on Agentic Software Engineering (SE 3.0), August 10, 2026, Jeju, Korea}
\acmISBN{}

\usepackage{tabularx}
\usepackage[most]{tcolorbox}

\begin{document}

\title{Trust but Verify? Uncovering the Security Debt of Autonomous Coding Agents }

\author{A H M Nazmus Sakib}
\affiliation{%
  \institution{University of Texas at San Antonio}
  \country{USA}}
\email{ahmnazmus.sakib@my.utsa.edu}

\author{Dipayan Banik}
\affiliation{
  \institution{Danovo Energy Solutions}
  \country{USA}}
\email{dipayan5175@gmail.com}

\author{Murtuza Jadliwala}
\affiliation{%
  \institution{University of Texas at San Antonio}
  \country{USA}}
\email{murtuza.jadliwala@utsa.edu}

\renewcommand{\shortauthors}{Sakib et al.}

\begin{abstract}

The increasing adoption of autonomous coding agents accelerates software development but also introduces scoped security risks within high-impact file paths that can outpace traditional human review capacity. While prior research has primarily evaluated these systems in terms of functional correctness and productivity, this paper presents a large-scale empirical study using the AIDev dataset to systematically characterize security code smells in agent-generated pull requests (PRs). Through a combination of a validated LLM-as-a-judge framework and manual qualitative analysis, we identify and classify security misconfigurations across 16,112 file changes spanning 4,022 pull requests. Our results reveal that 38.9\% of agent-generated PRs contain at least one security smell, with supply chain integrity issues accounting for 82.3\% of all detected security smells. Furthermore, hard-coded credentials constitute 99.6\% of all critical-severity security smells. Crucially, we find that human collaborators are responsible for introducing 67.6\% of genuine leaked secrets within these agent-assisted workflows, while existing automated and human review processes fail to detect 81.1\% of these credentials prior to integration. These findings highlight substantial security risks in agent-assisted software development workflows and suggest a potential reduction in developer vigilance. They also underscore the urgent need for context-aware security guardrails implemented directly at the point of human-AI collaboration.

\end{abstract}



\maketitle

\section{Introduction}

The increasing adoption of Artificial Intelligence (AI) tools in software engineering has contributed to the emergence of Software Engineering 3.0 (SE 3.0), a paradigm characterized by AI systems that can perform a range of software development tasks with limited human intervention\cite{li2025riseaiteammatessoftware, Hassan2026SE3}. Systems such as GitHub Copilot\cite{github_copilot}, OpenAI Codex\cite{openai_codex}, Claude Code\cite{claude_code}, Cursor\cite{cursor}, and Devin\cite{devin} can execute multi-step development workflows, including task planning, code generation, testing, and the creation of Pull Requests (PRs) for integration into existing codebases. Unfortunately, this acceleration often compromises system trustworthiness and code quality\cite{NegriRibalta2024}: the scale of agent-generated PRs outpaces human review capacity, leaving maintainers fatigued and over-reliant on automated outputs\cite{pimenova2025goodvibrationsqualitativestudy,sabra2025assessingqualitysecurityaigenerated}.

The mismatch between code production velocity and reviewer capacity has demonstrably compromised repository-level security integrity; nascent empirical analyses, albeit derived from limited datasets, suggest that AI-generated code exhibits a vulnerability introduction rate approximately 2.7 times that observed in human-written code\cite{dockerCodingAgent,substackLLMsCoding}. However, current software engineering literature predominantly evaluates Large Language Model (LLM) code generation with benchmarks focused on functional correctness rather than the security posture of the specific files and paths agents modify\cite{chen2021evaluatinglargelanguagemodels, jimenez2024swebench}. Although emerging empirical studies have begun to characterize agentic pull requests by analyzing their general modification patterns\cite{ogenrwot2026how}, test generation behaviors\cite{haque2026autonomousagentscontributetest, yoshimoto2026testingaiagentsempirical}, and commit quality\cite{rahman2026tasklevelevaluationaiagents}, there remains a critical research gap with respect to the introduction of structural security flaws within the specific code paths that agents touch. Conducting a large-scale, scoped empirical study on real-world data is imperative to document and understand the distribution, types, and severity of security code smells introduced by the use of autonomous agents.

To address this gap, we present an empirical study leveraging the AIDev dataset: a large-scale repository of PRs generated by autonomous coding agents to systematically characterize their security implications. In this paper, we focus on 'security smells' (structural patterns that indicate potential security risks), rather than confirmed exploitable vulnerabilities, which require context-specific exploitability analysis. We utilize a combination of LLM-as-a-judge evaluation, and manual qualitative analysis to investigate the following Research Questions (RQs):

\textit{RQ1: Which security smells appear most in agentic PRs, and how do they distribute across categories and severity?}

\textit{RQ2: How many flagged hard-coded secrets are genuine, who introduced them, and are they identified during review before integration?}

This paper makes the following core contributions:
\begin{itemize}
\item A six-category taxonomy of security smells in agent-generated PRs, grounded in OWASP secure-coding guidance, the Center for Internet Security Benchmarks, and GitHub hardening documentation.
\item An LLM-as-a-judge detection pipeline that categorizes security smells in agentic diffs with quantized open-source models, validated against a human-annotated gold standard using precision, recall, F1, and Cohen's kappa.
\item A large-scale empirical characterization of the prevalence, category distribution, and severity of security smells across agentic PRs in the AIDev dataset\cite{li2025riseaiteammatessoftware}.
\item A qualitative analysis of the most critical smells that traces each genuine credential to its committer, the agent or the human collaborator, and assesses whether automated bots and human reviewers intercept it before integration.

\end{itemize}

\begin{figure*}[!t]
    \centering
    \includegraphics[width=\textwidth]{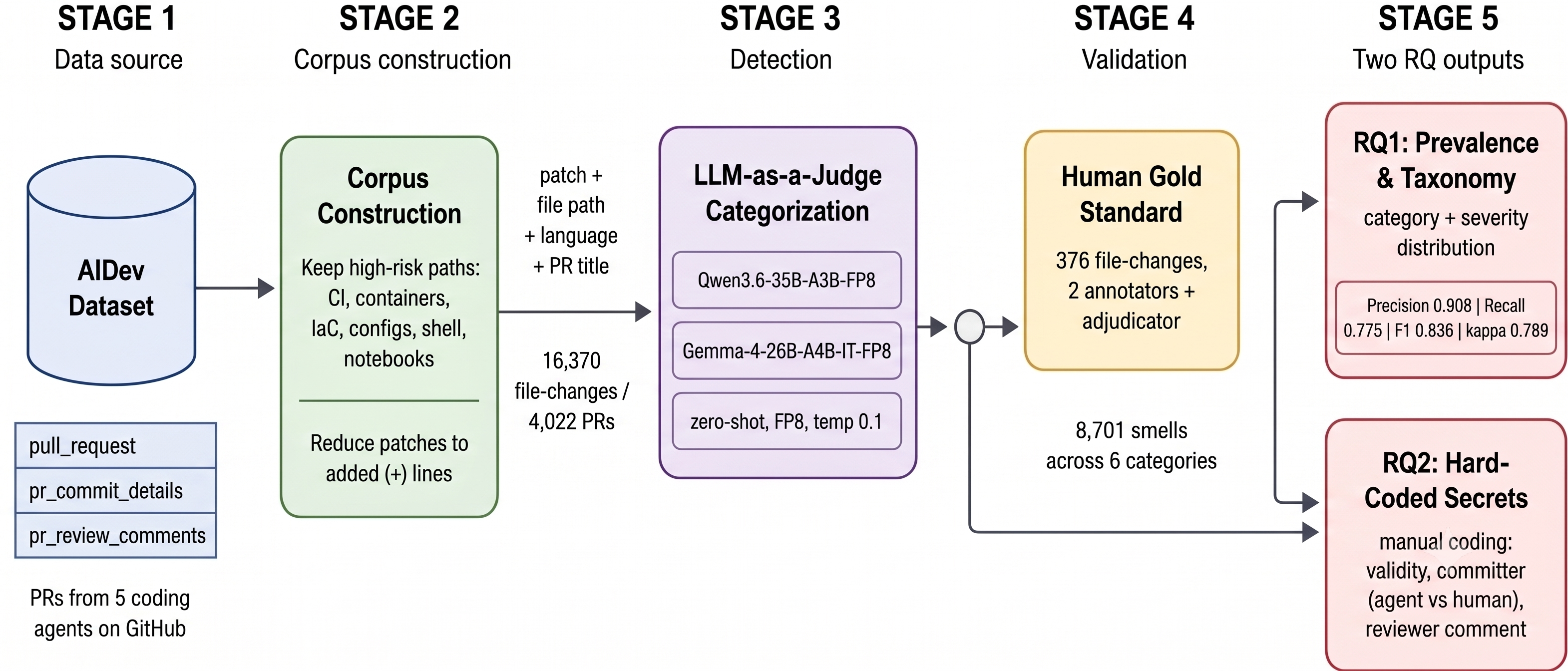}
    \caption{Architecture of the proposed methodology.}
    \label{fig:methodology}
\end{figure*}

\section{Methodology}
We use AIDev, a public dataset of PRs from five coding agents  on GitHub. We draw on three tables: \textit{pull\_request} holds per-request metadata, \textit{pr\_commit\_details} holds one row per changed file per commit with the unified-diff patch, and \textit{pr\_review\_comments} holds inline review comments. Our methodology is shown in Figure \ref{fig:methodology}.

\subsection{Detection and Quantification of Security Smells}
\textbf{Corpus construction.} We keep files whose path matches a high-risk pattern set covering Continuous Integration (CI) definitions, container files, infrastructure-as-code, secret-bearing files, configuration files, shell scripts, and computational notebooks. We exclude lock files, minified assets, vendor directories, and binary formats. We assign each retained file to the file categories to support stratified analysis. We then reduce each patch to the lines added, retaining only \textbf{+} lines and discarding diff metadata and removed lines, so that every counted smell reflects introduced content. The corpus contains 16,370 file changes across 4,064 PRs. Because we filter for high-risk patterns and analyze only added lines, this work constitutes a scoped security-smell analysis rather than a comprehensive repository audit.

\textbf{Taxonomy.} We define six smell categories, grounded in the OWASP guidance, the Center for Internet Security Benchmarks, and the GitHub hardening documentation\cite{owasp_secure_coding}, shown in Table \ref{tab:security_categories}.

\begin{table}
\centering
\caption{Security Smell Categories and Scope}
\label{tab:security_categories}
\small
\begin{tabularx}{\columnwidth}{@{}l X@{}}
\hline
\textbf{Category} & \textbf{Scope} \\
\hline
\texttt{secrets\_identity} &
Hard-coded keys, tokens, private keys, and credential-bearing connection strings. \\
\texttt{cleartext\_transport} &
Cleartext endpoints, disabled TLS, and weak ciphers. \\
\texttt{over\_privilege\_execution} &
Root containers, world-writable files, sudo, shell installers, broad CI scopes. \\
\texttt{permissive\_network} &
All-interface binds, open CIDRs, and public-access flags. \\
\texttt{misconfig\_hardening} &
Debug modes, logged secrets, and disabled encryption. \\
\texttt{supply\_chain\_integrity} &
Mutable action/image tags and unpinned global installs. \\
\hline
\end{tabularx}
\end{table}

\textbf{Detection.} We label the corpus with a large language model acting as a judge. We prompt two open-source models, Qwen3.6-35B-A3B-FP8\cite{qwen36_35b_a3b_fp8_2026} and 
Gemma-4-26B-A4B-IT-FP8\cite{gemma4_26b_a4b_it_fp8_2026}, with zero-shot prompting, at FP8 quantization through a local end-point at sampling temperature 0.1 (see Appendix \ref{subsection: llm}). The judge performs independent categorization over the whole corpus: for each file change, it receives the patch, file path, language, and PR title. The judge marks the file clean or flags each smell with one of the six categories, a line number and snippet, a severity on a three-level scale, and a one-sentence rationale. 

\textbf{Reliability and Validation.} We merge the runs of the models as a union, marking a file dirty when at least one run flags it, and deduplicating  smells on category, line number, and line snippet. We report inter-rater agreement with Cohen's Kappa and validate the judge against a random sample of 376 file changes with 95\% Confidence Level and 5\% Margin of Error. Two authors independently assign a smell label to each file change, and a third author adjudicates disagreements to a gold label matched to the judge. We report precision, recall, the F1 score, and Cohen's kappa against the gold label, together with inter-annotator agreement.

\textbf{Stratified Analysis.} To examine which factors drive smell prevalence, we disaggregate the flagging results along four dimensions: agent type, repository primary language, file category and PR size. We take the agent and the repository identifier from the pull request table and the primary language from the repository table. We assign the file category during corpus construction. We compute the PR size as the sum of changed lines over the distinct commits of a PR and bin it into five tiers: XS (1-9), S (10-49), M (50-199), L (200-999), and XL (1000+). We report a file-level flagging rate as the share of files marked dirty, a PR-level flagging rate as the share of PRs with at least one dirty file, and the mean smells per file.

\subsection{Validating and Reviewing Hard-Coded Secrets}

A hard-coded secret is the highest-impact security smell, because a live credential committed to a public repository is exploitable the moment it is pushed. RQ2 examines the secrets\_identity category in depth. We determine whether the secret is genuine, who committed it, and whether the existing reviewers catch it or not. We pay particular attention to automated security bots, since they are the layer expected to intercept a leaked credential before it causes harm.
We extract the complete set of smells that the categorization judge flags under secrets\_identity and manually examine each one. We carry no judge label into this stage so that the coders judge each secret on its own evidence.
Two authors independently code each secret in two dimensions. Validity records whether the flagged value is a genuine credential rather than a placeholder, an example, or a misread. Reviewer detection records whether an automated security bot commented on the secret. The coders inspect the live repository, the pull-request discussion, and the commit history that follows the introducing commit.
\section{Results}
To support reproducibility, we provide a replication package containing the data, scripts, and documentation required to reproduce the analyses reported in this paper. The replication package is available in the following link\cite{Kddworkshop2026:dataset}.

\subsection{RQ1: Distribution of Security Smells}

\begin{figure}[t]
  \centering
  \includegraphics[width=0.9\linewidth]{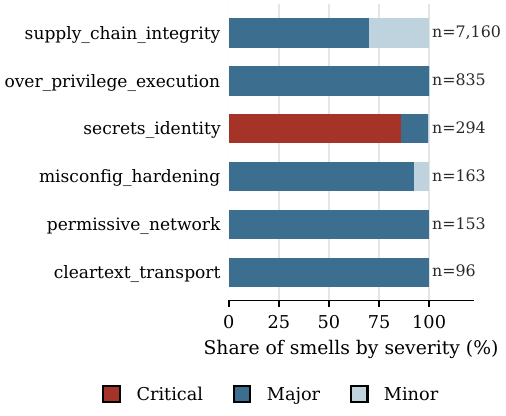}
  \caption{Severity composition of smells by category}
  \label{fig:smell_severity}
\end{figure}

The judge returned valid labels for 16,112 of the 16,370 file changes, which span 4,022 PRs. The judge failed to return parseable output for 258 file changes. The judge flagged 3,807 file changes as containing at least one smell, a rate of 23.6\%, and recorded 8,701 distinct smells after deduplication. At the PR level, 1,563 of 4,022 requests contained at least one smell, a rate of 38.9\%, and the remaining 2,459 requests stayed clean. The six categories differed by an order of magnitude in frequency, as Table \ref{tab:security_smells_distribution} reports. As shown in Figure \ref {fig:smell_severity}, the supply\_chain\_integrity category alone accounted for 7,160 of the 8,701 smells, a share of 82.3\%, followed by over\_privilege\_execution at 9.6\% and secrets\_identity at 3.4\%. The severity distribution placed 6,292 smells at the major level (72.3\%), 2,156 at the minor level (24.7\%), and 253 at the critical level (3.0\%). Critical smells concentrated in one category: 252 of the 253 critical smells, a share of 99.6\%, fell under secrets\_identity. The judge thus reserved its highest severity almost entirely for hard-coded credentials. The human validation set comprised 376 file changes where the two coders manually labeled each item and reached a Cohen's kappa of 0.929. The judge reached a precision of 0.908, a recall of 0.775, an F1 score of 0.836, and a Cohen's kappa of 0.789 against the gold labels. The judge therefore identified smells at high precision while missing 22.5\% of smells.

\begin{figure*}[t]
  \centering
  \includegraphics[width=0.8\linewidth]{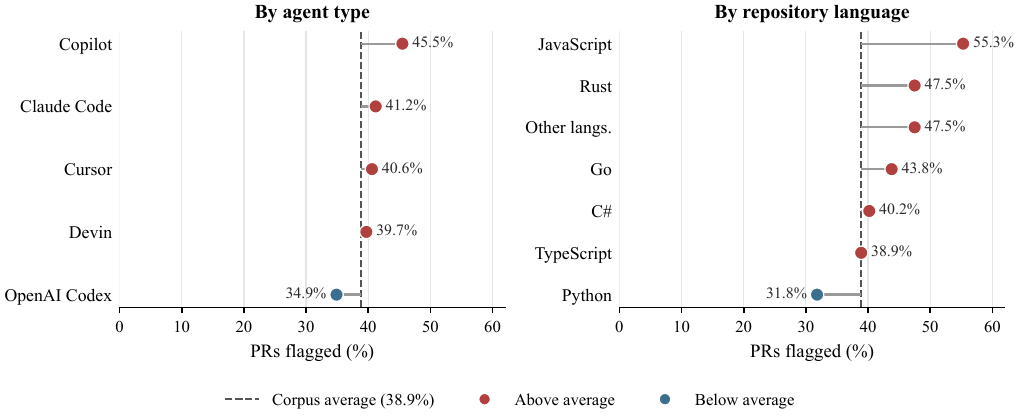}
  \caption{PR-level smell-flagging rate by agent type and by repository primary language, relative to the corpus average of 38.9\%.}
  \label{fig:lollipop_agent_language}
\end{figure*}

To identify the factors that drive smell prevalence, we disaggregated the PR-level flagging rate by agent, repository primary language, file category and PR size. Figure \ref{fig:lollipop_agent_language} compares the flagging rate against the 38.9\% corpus average by agent and language. Copilot PRs carried a smell most often at 45.5\%, and OpenAI Codex PRs least often at 34.9\%, yet Codex's flagged files carried the most smells per file at 0.62. JavaScript repositories showed the highest prevalence at 55.3\% and Python the lowest among the major languages at 31.8\%, a spread of 23.5 percentage points. TypeScript sat at the corpus average of 38.9\%.

\begin{figure}[t]
  \centering
  \includegraphics[width=0.9\linewidth]{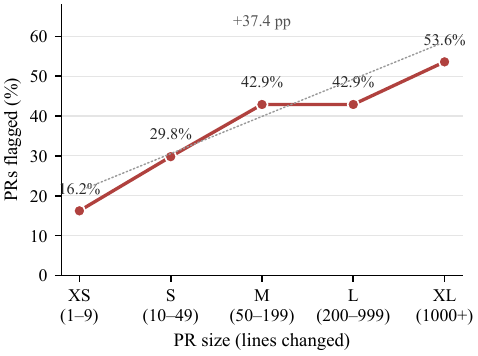}
  \caption{PR-level smell-flagging rate by PR size, showing an upwards trend with the amount of code changed.}
  \label{fig:line_prsize}
\end{figure}

Table \ref{tab:strata_filecat} disaggregates the file-level rate by file category. GitHub Actions workflows and Docker files together held 87.6\% of all smells, consistent with the concentration of \texttt{supply\_chain\_integrity} smells. The application configuration files remained clean at a 2.2\% dirty rate. Figure \ref{fig:line_prsize} groups the PRs by the total lines changed and shows the flagging rate increasing with the size of the PR, from 16.2\% for the smallest PRs (1-9 lines) to 53.6\% for the largest (1000+ lines), a spread of 37.4 percentage points.

\begin{tcolorbox}[
  title=Key Finding,
  colback=gray!5,
  colframe=black,
  fonttitle=\bfseries,
  sharp corners
]

 Supply\_chain\_integrity smell dominates at 82.3\% of the 8,701 smells, and hard-coded secrets form 99.6\% of the 253 critical smells. Prevalence rises with development activity from 16.2\% to 53.6\% of PRs by size, 87.6\% of smells in GitHub Actions and Docker files, highest for Copilot (45.5\%) and JavaScript repositories (55.3\%). The judge matched the human gold labels at 0.908 precision and 0.836 F1 score.
\end{tcolorbox}

\begin{table}[ht]
\centering
\caption{Distribution of Smells by Category and Severity}
\label{tab:security_smells_distribution}
\begin{tabularx}{\linewidth}{p{0.42\linewidth} *{5}{>{\raggedleft\arraybackslash}X}}
\hline
\textbf{Category} & \textbf{Smells} & \textbf{Critical} & \textbf{Major} & \textbf{Minor} \\
\hline
\texttt{supply\_chain\_integrity}   & 7,160 &  0   & 5,018 & 2,142 \\
\texttt{over\_privilege\_execution} & 835    & 1   & 833   & 1     \\
\texttt{secrets\_identity}          & 294     & 252 & 41    & 1     \\
\texttt{misconfig\_hardening}       & 163     & 0   & 151   & 12    \\
\texttt{permissive\_network}        & 153     & 0   & 153   & 0     \\
\texttt{cleartext\_transport}       & 96     & 0   & 96    & 0     \\
\hline
\end{tabularx}
\end{table}

\begin{table}[t]
\centering
\caption{Smell-Flagging Rate by File Category (file-level)}
\label{tab:strata_filecat}
\begin{tabularx}{\linewidth}{p{0.36\linewidth} *{3}{>{\raggedleft\arraybackslash}X}}
\hline
\textbf{Category} & \textbf{Files} & \textbf{\% Dirty} & \textbf{Smells} \\
\hline
GH Actions workflows & 7,054 & 36.3\% & 6,577 \\
Docker                    & 1,775 & 36.4\% & 1,058 \\
Other CI                  & 89    & 22.5\% & 40    \\
Shell scripts             & 2,142 & 12.0\% & 465   \\
Terraform                 & 374   & 11.5\% & 76    \\
K8s/Helm                  & 636   & 11.2\% & 119   \\
Notebooks                 & 793   & 10.1\% & 123   \\
App config                & 1,058 & 2.2\%  & 30    \\
\hline
\end{tabularx}
\end{table}

\subsection{RQ2: Genuine Secrets and Review Outcomes}

The coders assigned a validity label to 272 of the 294 secret smells. The remaining 22 could not be labeled because the PRs no longer exist. The coders confirmed 74 genuine credentials, a rate of 27.2\% of the labeled set. The commit author of each genuine secret fell into one of two classes. A human author committed 50 of the 74 genuine secrets, a rate of 67.6\%, and an AI agent committed the remaining 24. The human collaborator therefore introduces genuine secrets into agent pull requests more often than the agent does.

\begin{figure}[H]
    \centering
    \includegraphics[width=\linewidth]{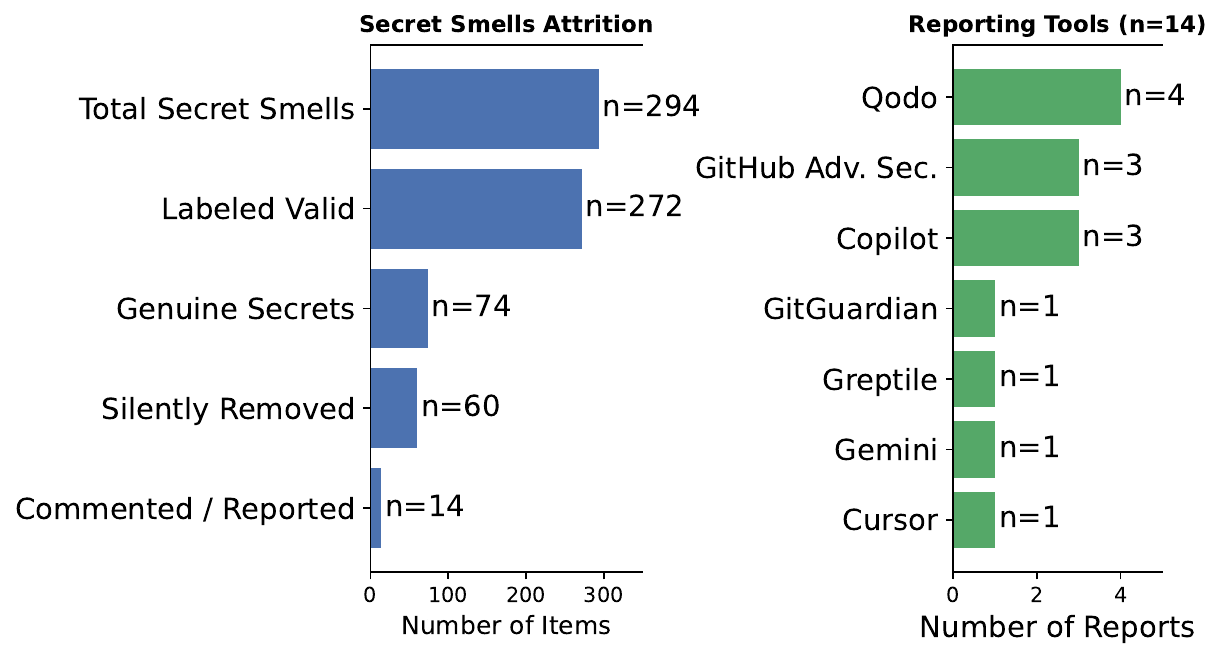}
    \caption{Summary of RQ2 Investigation Outcomes}
    \label{fig:rq2_attr}
\end{figure}

A security bot or a human reviewer commented on the secret in 14 of the 74 genuine cases, a rate of 18.9\%, and the reviewers removed the other 60 secrets without a comment. The commenting actors comprised seven distinct tools and reviewers: GitHub Advanced Security, GitGuardian, Qodo, Copilot, Greptile, Gemini Code Assist, and Cursor bot. A summary of these results is depicted in Figure \ref{fig:rq2_attr}.

\begin{tcolorbox}[
  title=Key Finding,
  colback=gray!5,
  colframe=black,
  fonttitle=\bfseries,
  sharp corners
]
Coders confirmed 27.2\% of labeled secret smells as genuine credentials. Human collaborators committed 67.6\% of them, and 81.1\% reached integration without a review comment from a bot or human.
\end{tcolorbox}

\section{Threats to Validity}

The study relies on a manually annotated gold standard of 376 file changes, making results susceptible to annotation bias and human error. Similar risks apply to the manual validation of hard-coded secrets in repositories and commit histories. To reduce these threats, we used independent double-blind coding and third-author adjudication.

The use of quantized LLMs (Qwen3.6 and Gemma-4) introduces limitations such as non-determinism and incomplete detection. With a recall of 0.775, the LLM judge missed 22.5\% of security smells, meaning reported frequencies likely underestimate the true prevalence of security smells.

By intentionally restricting the analyzed corpus to high-risk file paths and focusing exclusively on added lines, the findings characterize specific security smells introduced within targeted boundaries rather than providing a comprehensive audit of total repository security health. Security smells or misconfigurations present in excluded files or introduced through complex interactions across multiple existing files are not captured in this analysis.

The findings are based on the AIDev dataset, which focuses on selected coding agents in open-source GitHub repositories. As a result, the observed security smells and review patterns may not generalize to other agents, enterprise environments, or development platforms.

\section{Related Work}
The advent of autonomous coding agents has prompted extensive empirical investigation into the differences between AI-generated and human-authored PRs. Ogenrwot\cite{ogenrwot2026how} found substantial differences in commit counts, merge outcomes, lifecycle dynamics, and description-to-diff alignment compared to human baselines. Yoshioka et al.\cite{yoshioka2026letsmakepullrequest} identified that while submitter attributes dominate merge success in both groups, review-related features exert contrasting effects on agentic PRs. This builds upon Watanabe et al.\cite{watanabe2026useagenticcodingempirical}, who established the distinct architectural footprint of AI contributors. Rahman et al.\cite{rahman2026tasklevelevaluationaiagents} conducted task-level evaluations demonstrating that while certain models achieve high PR acceptance rates, others trigger significantly higher volumes of human and automated review discussions. Haque et al.\cite{haque2026autonomousagentscontributetest} and Yoshimoto et al.\cite{yoshimoto2026testingaiagentsempirical} observed that agent-generated test methods present distinct structural patterns. The human-AI interaction layer is equally critical; Haider et al.\cite{haider2026understandingdominantthemesreviewing} taxonomized reviewer responses to AI-authored code, revealing that inline comments predominantly address logical and functional correctness rather than security posture.

While AI applications provide a foundation for open-source software sustainability\cite{karim2026artificialintelligenceopensource}, system trustworthiness remains an underexplored bottleneck\cite{roychoudhury2025agenticaisoftwareengineers}. Prior security literature on PRs has primarily focused on human-authored vulnerabilities or automated patching mechanisms utilizing datasets like TREEVUL\cite{treevul} or architectures like GraphSPD\cite{Wang2023GraphSPDGS}. Wang et al.\cite{felixwaterloo} recently addressed AI-specific security dynamics by examining automated versus human security patching patterns within the AIDev dataset. However, while these studies analyze patching efficacy, functional correctness, or general modification metrics, our work distinctly isolates the introduction of structural security smells (e.g., hard-coded secrets, supply chain misconfigurations) during autonomous generation.
\section{Conclusion and Future Work}

This study provides a large-scale empirical analysis of security code smells introduced within autonomous agent-generated pull requests. By analyzing the AIDev dataset using a combination of validated LLM-as-a-judge methodology and manual qualitative analysis, we established that agentic coding workflows carry significant security debt. Code smells were present in 38.9\% of agent-authored PRs, with supply chain integrity misconfigurations constituting the vast majority (82.3\%) of identified security smells. Furthermore, our investigation into critical-severity smells revealed that 99.6\% are hard-coded credentials. Crucially, our findings show that human collaborators were mostly  responsible, injecting 67.6\% of genuine leaked secrets within agentic PRs. Compounding this risk, the current safety net of automated security bots and human reviewers failed to intercept 81.1\% of these leaked credentials prior to integration.

These results reveal a critical operational insight: the velocity of agentic code generation outpaces and compromises traditional human-in-the-loop review pipelines. The high rate of human-introduced secrets suggests potential review fatigue or cognitive offloading, though further behavioral studies are required to determine why developers bypass standard security hygiene when interacting with autonomous workflows.

Future research must focus on mitigating developer cognitive load by designing context-aware security interfaces that evaluate agentic workflows without exacerbating review fatigue. Investigations should also explore novel mechanisms for actively enforcing security hygiene at the point of human-AI collaboration, specifically targeting preventative measures against the initial injection of live credentials. Finally, expanding this empirical analysis to enterprise environments is necessary to determine if proprietary guardrails alter security smell distributions or better sustain developer vigilance.

\bibliographystyle{ACM-Reference-Format}
\bibliography{sample-base}

\clearpage
\appendix

\section{Appendix}

\subsection{LLM Setup Information}
\label{subsection: llm}
\begin{figure}[!htbp]
    \centering
    \includegraphics[width=1\linewidth]{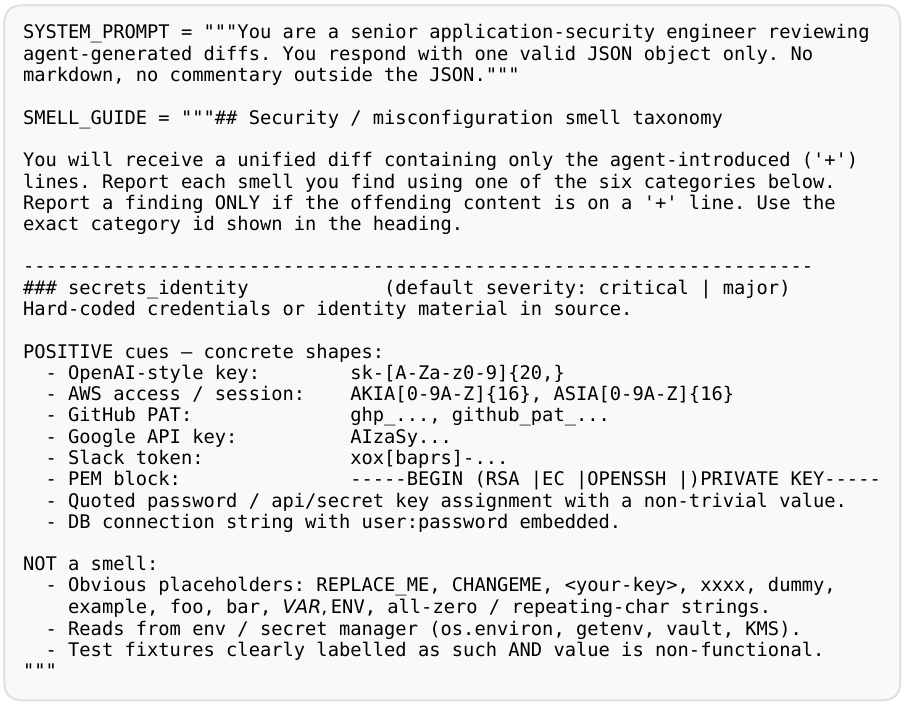}
    \caption{Prompt for LLM-as-a-judge}
    \label{fig:prompt}
\end{figure}

\begin{figure}[!htbp]
    \centering
    \includegraphics[width=1\linewidth]{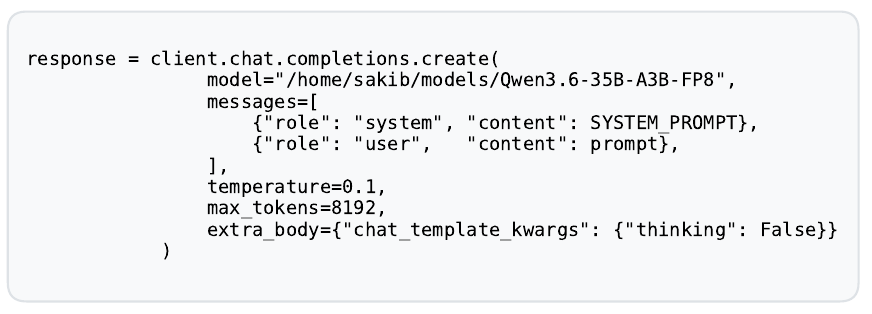}
    \caption{Sample Response sent to LLM}
    \label{fig:response}
\end{figure}

A sample of the prompt provided to the LLM-as-a-judge system is shown in Figure \ref{fig:prompt}. A sample response sent to LLM is shown in Figure \ref{fig:response}. 

\begin{figure}[!htbp]
    \centering
    \includegraphics[width=1\linewidth]{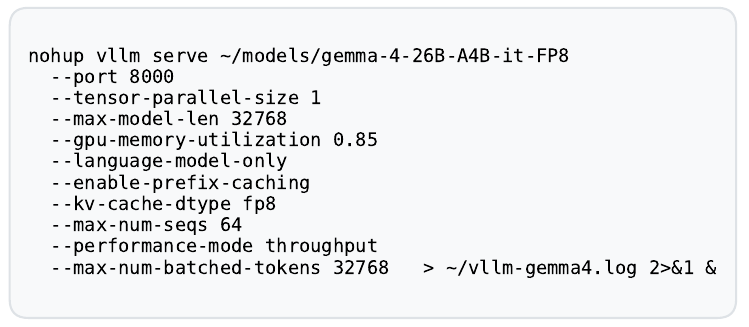}
    \caption{Sample LLM Server Setup Command}
    \label{fig:vllm}
\end{figure}

The LLMs were hosted locally using VLLM backend with the command shown in Figure \ref{fig:vllm}. An NVIDIA RTX PRO 6000 Blackwell Workstation Edition Graphics Card with 96GB VRAM was used to host the LLMs.

\subsection{Sample Secrets Found During Manual Verification}

\begin{figure}[!htbp]
    \centering
    \includegraphics[width=0.9\linewidth]{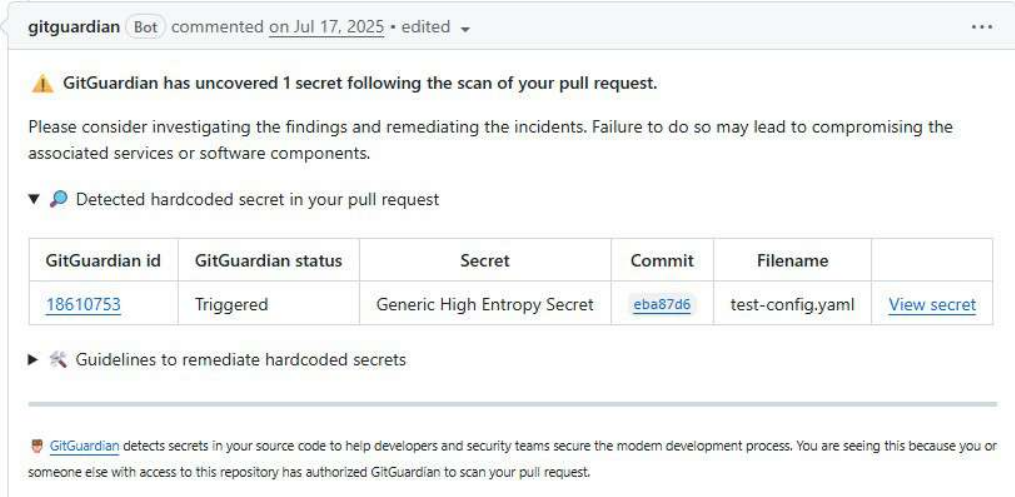}
    \caption{Sample Secret 1}
    \label{fig:ss1}
\end{figure}

\begin{figure}[!htbp]
    \centering
    \includegraphics[width=0.9\linewidth]{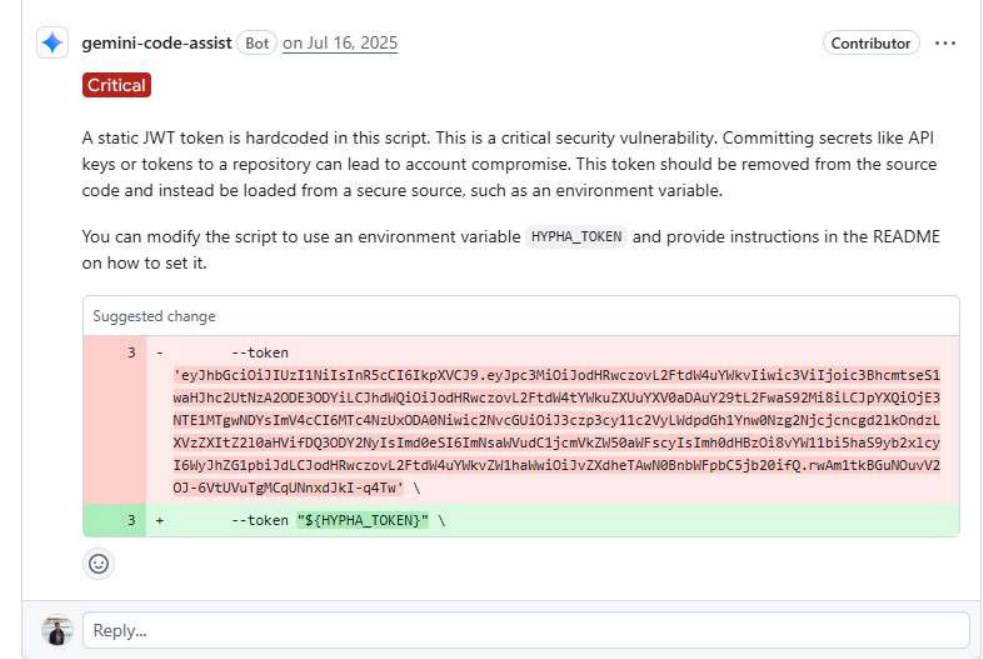}
    \caption{Sample Secret 2}
    \label{fig:ss2}
\end{figure}

Figures \ref{fig:ss1} and \ref{fig:ss2} show two sample secrets found while manually verifying the results produced by the LLM-as-a-judge system.

\end{document}